\documentclass[proceedings]{JHEP3}
\usepackage[math]{iwona}
\usepackage{amsmath,amssymb,amsthm,mathrsfs,amsfonts,dsfont}
\usepackage{epsfig,multicol}
\usepackage{bbold}
\usepackage{xcolor}

\newbox\mybox

\newcommand\fverb{\setbox\mybox=\hbox\bgroup\verb}
\newcommand\fverbdo{\egroup\medskip\noindent\fbox{\unhbox\mybox}\ }
\newcommand\fverbit{\egroup\item[\fbox{\unhbox\mybox}]}
\conference{The evolution of spectral data for nonlinear Klein-Gordon models}

\abstract{We investigate the effect of the breaking of integrability in the integrals of motion of a sine-Gordon-like system. The class of quasi-integrable models, discussed in the literature, inherits some of the integrable properties they are associated with. Our strategy, to investigate the problem through a deformation of the so-called inverse scattering method, has proven to be useful in the discussion of generic nonlinear Klein-Gordon potentials, as well as in particular cases presented here.}

\title{The evolution of spectral data for nonlinear Klein-Gordon models}

\author{P. H. S. Palheta, P. E. G. Assis, T. M. N. Gon\c{c}alves \\
Universidade Federal de Catal\~ao, Goi\'as, Brazil\\
E-mail: paulo.assis@ufcat.edu.br}

\begin{document}

\setcounter{equation}{0}

\section{Introduction}\label{secintro}\qquad

Integrable systems possess remarkable mathematical properties, such as the existence of an infinite number of conserved quantities, allowing for the solvability of the associated equations of motion. 
Their study not only deepens our understanding of the underlying fundamental structures, it also provides valuable insights into the behavior of complex systems, ranging from classical mechanics to quantum field theory
\cite{moser2011aspects,faddeev2007hamiltonian,babelon2003introduction,korepin1997quantum,drazin1989solitons}.
It is no surprise that integrable systems  and solitons play a crucial role in theoretical and applied mathematics, as well as in various areas of science, possessing an interdisciplinary nature \cite{remoissenet2013waves}. 
They arise in diverse contexts, including nonlinear waves, statistical mechanics, and quantum information theory, and their significance extends beyond pure mathematics to interdisciplinary research areas, from nonlinear physics to engineering and biology \cite{takeno2013dynamical}. 
In the realm of nonlinear optics, solitons are identified in the transmission of information in optical fibers, enabling long-distance communication with minimal distortion \cite{maimistov2001completely}.
Additionally, for applications in fluid dynamics, soliton-like solutions of integrable models describe phenomena, such as waves in shallow water \cite{korteweg1895xli, peterson2003soliton}  or  in traffic flow \cite{hattam2017kdv}. 
Furthermore, in condensed matter physics, integrable systems provide insights into the behavior of quantum many-body systems, with applications ranging from the study of quantum spin chains \cite{klumper2008integrability} to the description of electron transport in one-dimensional conductors \cite{jesko2020transport} and polymers \cite{heeger1988solitons}.
Finally, in theoretical biology, soliton-like solutions have been proposed to describe the propagation of electrical signals along nerve fibers \cite{heimburg2005soliton} and the nonlinear dynamics associated to DNA \cite{yakushevich2002nonlinear}.

Lax pairs and the zero curvature representation are powerful mathematical tools used in the study of integrable systems, particularly in the context of nonlinear partial differential equations (PDEs) and soliton theory \cite{ablowitz1991solitons,novikov1984theory,ablowitz1981solitons,kaup1975method,ablowitz1974inverse,ablowitz1973method,rubinstein1970sine}.
This formalism provides a systematic way to generate infinitely many conserved quantities associated with the integrable system, leading to the complete integrability of the equations, ensuring the existence of an infinite number of symmetries and the determination of soliton solutions \cite{zakharov1974scheme,ablowitz1974inverse,gardner1967method}. 
It provides a framework for solving integrable systems of partial differential equations  by the inverse scattering method by mapping solutions of the given nonlinear differential to those of the associated linear system. By analyzing the scattering data and employing techniques from complex analysis and spectral theory, one can then recover the solutions of the original equation through an inverse transformation
\cite{kay1956determinationIII,kay1956determinationII,kay1955determinationI,krein1955determination,marchenko1955reconstruction,jost1952construction,gel1951determination}.

Nevertheless, despite their mathematical elegance and remarkable properties, integrable systems constitute only a small fraction of the vast landscape of nonlinear equations that arise in various fields of physics, engineering, and applied mathematics. While integrable systems possess desirable features, such as soliton solutions and infinitely many conserved quantities, many real-world phenomena cannot be accurately described solely within the framework of integrable models. Therefore, it is crucial to go beyond integrable systems and consider quasi-integrable systems \cite{ferreira2012concept,ferreira2012attempts,ferreira2011concept,ferreira2011some,bazeia2008classical}, which represent a larger and more realistic class of problems encountered in practice. Quasi-integrable systems exhibit some - but not all - of the characteristics of integrability, allowing for more flexible and nuanced descriptions of complex physical phenomena \cite{blas2022modified,abhinav2022non,ter2019quasi}. 
By extending the current theoretical framework to encompass quasi-integrable systems, one can hope to capture the richness and diversity of real-world dynamics, proposing a path towards a more comprehensive understanding of the complex behavior observed in nature.

This work is organized as follows. In section \ref{secQuasiInt}, we present possible representations of the nonlinear Klein-Gordon under the influence of a generic potential, in terms of Lax pairs whose curvature does not vanish identically. Considering the vanishing of the curvature for some circumstances, asymptotic solutions of the linear problem are presented. Then, in section \ref{secSpectralData} we introduce the quasi-integrable spectral data of this system, generalizing the inverse scattering scheme. It is possible to determine, in section \ref{secTimeEvol}, the time evolution of these objects to conclude that for certain boundary conditions these objects are conserved in time. Finally, in section \ref{secStudyPotential},  we numerically verify their constancy by considering various forms of potentials, including modifications of the sine-Gordon model and the $\phi^4$ theory. We summarize our findings in section \ref{secConclusion}. 

\section{Quasi-integrable representation of the nonlinear Klein-Gordon equation} \label{secQuasiInt}\qquad

The simplest example of an integrable system is perhaps given by the equation \\
$$
\ddot{q}=-\frac{1}{m} V'(q),
$$ representing the dynamics of a particle of mass $m$ under the influence of a one-dimensional conservative potential $V(q)$,
which can be represented by the Lax equation \cite{krishnaswami2020introduction} 
\begin{equation}\label{nonlinearKleinGordon}
\frac{\partial L}{\partial t} = [M,L],
\end{equation}
with
\begin{eqnarray}
	L
	~=~
	\left(
	\begin{array}{cc}
		p & \sqrt{2m \, V(q)} \\
		\sqrt{2m \, V(q)} & - p \\
	\end{array}
	\right)
	\quad
	\textrm{and}
	\quad
	M
	~=~
	\left(
	\begin{array}{cc}
		0 & 
		-\frac{1}{4} \,
		\frac{\frac{2}{m}\,V'(q)}{\sqrt{\frac{2}{m}\,V(q)}}
		\\
		+\frac{1}{4} \,
		\frac{\frac{2}{m}\,V'(q)}{\sqrt{\frac{2}{m}\,V(q)}}& 0 
	\end{array}
	\right).
	\qquad\qquad
	\nonumber
\end{eqnarray}
The conserved quantities associated to \eqref{nonlinearKleinGordon} can be shown to be 
$
	\text{Tr}\; L^{2(k+1)}
	~=~
	\big(2m \, H \big)^{k+1},
$
essentially just powers of the total energy $H=\frac{p^2}{2m} +V(q)$, which can then be used to solve the problem by quadrature. 

In a similar fashion as for the one-dimensional particle just discussed, one can look now for a curvature condition for a generic relativistic classical field, starting with a free massless $1+1$ field $\phi(x,t)$, described by the D'Alembert wave equation, and introduce an external potential, $\mathcal{V}(\phi)$, 
\begin{eqnarray}
	\Box \phi + \mathcal{V}'(\phi) ~=~ 0.
\end{eqnarray}
Thus, one obtains an equation that generalizes the Klein-Gordon equation, for when the potential is
$\mathcal{V}(\phi) = \frac{\mu^2}{2} \phi^2$, and that includes the integrable sine-Gordon model, for when it is described by
$\mathcal{V}(\phi) = \frac{\mu^2}{\beta^2} \left( 1 - \cos \beta\phi \right)$ so that
$\Box \phi + \frac{\mu^2}{\beta} \sin \beta\phi ~=~ 0$ is satisfied.
The stable vacua have configurations $\phi_\pm$ for which the potential is a constant and no force is felt on the system so $\mathcal{V}(\phi_\pm) = \mathcal{V}_\pm$ and $\mathcal{V}'(\phi_\pm) = 0$; for the sine-Gordon case, one would have stable solutions located at the minima $\phi_\pm = \frac{2N_\pm \, \pi}{\beta}$ with $N_\pm \in \mathds{N}$.

Starting from a standard form of a Lax pair associated with the sine-Gordon model, we express it in terms of a generic potential $\mathcal{V}(\phi)$ so, as can be verified, the deformed potentials in the Lax pair can be written as
\begin{eqnarray}\label{standardLaxPairSGMt}
	A_t(\lambda;x,t)
	&=&
	\left(
	\begin{array}{cc}
		-\frac{i \beta}{4}  \partial_x\phi & 
		-\frac{i}{4}  
		\left(\frac{\left(\lambda +\frac{1}{\lambda }\right)
			\mathcal{V}'(\phi)}{\sqrt{2} \sqrt{\mathcal{V}(\phi)}}
		-\frac{i \beta  \left(\lambda -\frac{1}{\lambda }\right)\sqrt{\mathcal{V}(\phi)}}{\sqrt{2}}\right) 
		\\
		-\frac{i}{4} 
		\left(\frac{\left(\lambda +\frac{1}{\lambda }\right) \mathcal{V}'(\phi)}{\sqrt{2} \sqrt{\mathcal{V}(\phi)}}
		+\frac{i \beta  \left(\lambda -\frac{1}{\lambda }\right) \sqrt{\mathcal{V}(\phi)}}{\sqrt{2}}\right) &
		\frac{i \beta}{4}  \partial_x\phi 
	\end{array}
	\right),
	\qquad
	\\\label{standardLaxPairSGMx}
	A_x(\lambda;x,t)
	&=&
	\left(
	\begin{array}{cc}
		-\frac{i \beta}{4}  \partial_t\phi & 
		-\frac{i}{4}  
		\left(\frac{\left(\lambda -\frac{1}{\lambda }\right)
			\mathcal{V}'(\phi)}{\sqrt{2} \sqrt{\mathcal{V}(\phi)}}
		-\frac{i \beta  \left(\lambda +\frac{1}{\lambda }\right)\sqrt{\mathcal{V}(\phi)}}{\sqrt{2}}\right) 
		\\
		-\frac{i}{4} 
		\left(\frac{\left(\lambda -\frac{1}{\lambda }\right) \mathcal{V}'(\phi)}{\sqrt{2} \sqrt{\mathcal{V}(\phi)}}
		+\frac{i \beta  \left(\lambda +\frac{1}{\lambda }\right) \sqrt{\mathcal{V}(\phi)}}{\sqrt{2}}\right) &
		\frac{i \beta}{4}  \partial_t\phi 
	\end{array}
	\right),
	\qquad
\end{eqnarray}
with choice of vacuum so that $\mathcal{V}(\phi)\ge 0$ and $\frac{\mathcal{V}'(\phi)}{\sqrt{\mathcal{V}(\phi)}}$ is finite. 
Indeed, the associate curvature,\footnote{A compatibility condition for the system composed of the two linear problems, $\big( \partial_x + A_x(\lambda;x,t) \big) \Psi(\lambda;x,t) =0$ and $\big( \partial_t + A_t(\lambda;x,t) \big) \Psi(\lambda;x,t)=0$.}
$F_{xt} ~=~ \partial_x A_t - \partial_t A_x + \left[ A_x, A_t \right]$, 
in terms of Pauli $\sigma$ matrices, reads
\begin{eqnarray}
	\frac{F_{xt}}{\frac{i \beta}{4}}
	~=~
	\big( \Box \phi + \mathcal{V}'(\phi) \big) \, \sigma_z
	+
	\left[ 
	\left(\lambda-\frac{1}{\lambda}\right) \partial_t \phi
	-
	\left(\lambda+\frac{1}{\lambda}\right) \partial_x \phi
	\right]
	\frac{\left[
		2 \mathcal{V}''(\phi)
		-\frac{\mathcal{V}'(\phi)^2}{\mathcal{V}(\phi)}
		+\beta^2 \mathcal{V}(\phi)
		\right]}{2\sqrt{2}\beta\,\sqrt{\mathcal{V}(\phi)}}
	\, 
	\sigma_x,
	\qquad\qquad
	\nonumber
\end{eqnarray}
which corresponds to a term that is equivalent to the equation of motion we are interested in, followed by an anomalous term which vanishes automatically if the potential is that of the sine-Gordon model.
However, asymptotically as the fields tend to a vacuum configuration and their derivatives tend to vanish, this anomaly disappears, allowing for the introduction of quasi-integrable systems. Nonetheless, it is also possible to have a vanishing curvature in the case where the fields travel with a particular speed and, interestingly, we note that the zero curvature condition may depend not only on the model - as it is the case for integrable systems - but also on the solution \cite{assis2016symmetry}. This opens up the possibility of constructing conserved charges by deformations of the inverse scattering method for the sine-Gordon equation.
To determine the solution as $x \to \infty$,  $\Psi^{\pm}(\lambda;x)=\left(\vec{\psi}_1^{\pm}(\lambda;x)\;\;\vec{\psi}_2^{\pm}(\lambda;x)\right)$, one needs to solve the linear problem 
\begin{eqnarray}\label{eqWithKs}
	\big( \partial_x + A_x^\pm(\lambda) \big) \, \Psi^\pm(\lambda;x) = 0
\quad\textrm{ with }\quad
	A_{x}^{\pm} = \begin{pmatrix}
		0 & -iK_{1}^{\pm} - K_{2}^{\pm}  \\
		-iK_{1}^{\pm} + K_{2}^{\pm}  & 0
	\end{pmatrix},
\end{eqnarray}
where
$
K_1^\pm =
\frac{1}{4}  
\frac{\left(\lambda -\frac{1}{\lambda }\right)
	\mathcal{V}'(\phi_\pm)}{\sqrt{2} \sqrt{\mathcal{V}(\phi_\pm)}} \textrm{ and }
K_2^\pm =
\frac{1}{4}\frac{\beta  \left(\lambda +\frac{1}{\lambda }\right)\sqrt{\mathcal{V}(\phi_\pm)}}{\sqrt{2}}.
$
Setting $k_{\pm}=\sqrt{{K_1^{\pm}}^2+{K_2^{\pm}}^2}$, it is possible to obtain the asymptotic solution,
\begin{eqnarray}
\Psi^{\pm}(\lambda;x) \label{psio}
=
\frac{1}{2}\begin{pmatrix}
e^{+ik_{\pm}x} & \frac{-K_1^{\pm}+iK_2^{\pm}}{k_{\pm}}e^{-ik_{\pm}x}\\
\frac{K_1^{\pm}+iK_2^{\pm}}{k_{\pm}}e^{+ik_{\pm}x} & e^{-ik_{\pm}x}
\end{pmatrix}.
\end{eqnarray}
At far away distances we only have incoming and outgoing waves, which are scattered by the interaction of the gauge potential. 
In a setting like this, the solutions we can construct are valid almost everywhere in space, except in a finite domain near the scattering, and from them 
one can clearly see that the nonlinear Klein-Gordon potential not only determines the momentum carried by the plane wave,
with wave vectors depending continuously on the spectral parameter,
but it can also introduce phase shifts.

The importance of these states is that they can be used in the asymptotic expansion of the general solution $\Psi(\lambda;x,t)$, whose time evolution allows for the reconstruction of the field dynamics when the curvature vanishes.
Moreover, if one imposes some conditions on the asymptotic behaviour of the physical potential $\mathcal{V}(\phi_+)=\mathcal{V}(\phi_-)$ and ${\mathcal{V}'(\phi_+)} = {\mathcal{V}'(\phi_-)}$, indicating that the potentials of interest must have degenerate vacua, with topological properties, then the gauge potentials, $A_\mu$, have to have the same asymptotic forms.

It is also worth mentioning that the choice for the gauge potentials is not unique, as other pairs can be constructed, such as
\begin{eqnarray}
	A_t(\lambda;x,t)
	&=&
	\left(
	\begin{array}{cc}
		-\frac{i \beta}{4} \left(\partial_t\phi-\partial_x\phi\right) & 
		-\frac{i \beta}{4} \mathcal{V}'(\phi)
		+\frac{1}{4} \left(\beta^2 \mathcal{V}(\phi)-\mu ^2\right)
		+\frac{1}{4\lambda} 
		\\
		\frac{i \beta}{4}  \lambda \mathcal{V}'(\phi)
		+\frac{1}{4} \lambda \left(\beta^2 \mathcal{V}(\phi)-\mu^2\right)
		+\frac{1}{4} & 
		\frac{i \beta}{4}  \left(\partial_t\phi-\partial_x\phi\right) \\
	\end{array}
	\right),
	\qquad
	\\
	A_x(\lambda;x,t)
	&=&
	\left(
	\begin{array}{cc}
		\frac{i \beta}{4} \left(\partial_t\phi-\partial_x\phi\right) & 
		-\frac{i \beta}{4} \mathcal{V}'(\phi)
		+\frac{1}{4} \left(\beta^2 \mathcal{V}(\phi)-\mu^2\right)
		-\frac{1}{4\lambda} 
		\\
		\frac{i \beta}{4} \lambda  \mathcal{V}'(\phi)
		+\frac{1}{4} \lambda \left(\beta^2 \mathcal{V}(\phi)-\mu^2\right)-\frac{1}{4} & 
		-\frac{i \beta}{4} \left(\partial_t\phi-\partial_x\phi\right) \\
	\end{array}
	\right),
	\qquad
\end{eqnarray}
with curvature vanishing again for the sine-Gordon potential, but also for other potentials, as long as the travelling wave configurations are of the form $\phi(x,t)=f(x+t)$, as shown below
\begin{eqnarray}
	\frac{F_{xt}}{\frac{i \beta}{4}}
	=
	-
	\left( \Box \phi + \mathcal{V}'(\phi) \right) \, f_0
	+
	\left( \mathcal{V}''(\phi) +  \beta^2 \mathcal{V}(\phi)-\mu ^2  \right) 
	\big(
	\partial_t \phi -\partial_x \phi
	\big)\, f_1,
	\qquad\qquad
	\nonumber
\end{eqnarray}
where we adopt $f_0$ and $f_1$ from the algebra defined in \cite{ferreira2012attempts}, i.e.
	\begin{eqnarray}
		f_{0}=\left(
		\begin{array}{cc}
			1 & 0 \\
			0 & -1
		\end{array}
		\right), ~~
		f_{1}=\left(
		\begin{array}{cc}
			0 & 1 \\
			-\lambda & 0
		\end{array}
		\right) ,~~
		b_{1}=\left(
		\begin{array}{cc}
			0 & 1 \\
			\lambda & 0
		\end{array}
		\right) ,~~ 
		b_{-1}=\left(
		\begin{array}{cc}
			0 & \frac{1}{\lambda} \\
			1 & 0
		\end{array}
		\right).
		\qquad\qquad
	\end{eqnarray}
The advantage of this last gauge, as the authors argue, is that the vacua potentials do not depend on any $f_{2n}= 2T_3^n$, so they live in an oscillator algebra - a commuting algebra if we are on a loop algebra, with no central extension.
For energy-shifted potentials, so that the minima positions, $\mathcal{V}'_\pm = 0$, satisfy $\mathcal{V}_\pm = 0$, and subsequently, $K_2^{\pm}$ vanishes and $K_1^{\pm}$ is finite, the Lax potentials are given by the commuting elements
$
	A_t^\pm 
	=
	- \frac{\mu^2}{4} \, b_1
	+ \frac{1}{4} \, b_{-1}$
	and
	$
	A_x^\pm 
	=
	- \frac{\mu^2}{4} \, b_1
	- \frac{1}{4} \, b_{-1}.  
$   
In fact, this choice of energy shift can be made in general, even for the previous gauge potentials, simplifying the asymptotic solution in Eq. (\ref{psio}).

\section{The spectral data in a quasi-integrable system}\label{secSpectralData}\qquad

Again, considering the linear auxiliary problem
\begin{eqnarray}
	\big( \partial_x + A_x(\lambda;x,t) \big) \, \vec{\psi}_i^{(J)}(\lambda;x,t) &=& 0
\end{eqnarray}
where $J=I,II$, so $\vec \psi_{1}^{(I)}(\lambda;x,t)$ and $\vec \psi_{2}^{(I)}(\lambda;x,t)$ denote two linearly independent solutions with a certain boundary condition to the left, whereas $\vec \psi_{1}^{(II)}(\lambda;x,t)$ and $\vec \psi_{2}^{(II)}(\lambda;x,t)$ are other linearly independent solutions with another boundary condition to the right.
The boundary conditions are given by asymptotic solutions of the vacuum linear equation
and these solutions can be related to each other according to
\begin{eqnarray}\label{abexpansion1}
	\vec \psi_{1}^{(I)}(\lambda;x,t)
	&=&
	a(\lambda;x,t) \, \vec \psi_{1}^{(II)}(\lambda;x,t)  
	+
	b(\lambda;x,t) \, \vec \psi_{2}^{(II)}(\lambda;x,t), 
	\\[8pt]\label{abexpansion}
	\vec \psi_{2}^{(I)}(\lambda;x,t)
	&=&
	\hat{b}(\lambda;x,t) \, \vec \psi_{1}^{(II)}(\lambda;x,t)  
	+
	\hat{a}(\lambda;x,t) \, \vec \psi_{2}^{(II)}(\lambda;x,t).
\end{eqnarray}
The expansion coefficients in \eqref{abexpansion1} and \eqref{abexpansion} can be related to the transmission and reflection coefficients through
\begin{eqnarray}
	\nonumber
	\tau(\lambda;x,t)
	=
	\frac{1}{a(\lambda;x,t)},
	\quad
	\hat{\tau}(\lambda;x,t)
	=
	\frac{1}{\hat{a}(\lambda;x,t)},
	\quad
	r(\lambda;x,t)
	=
	\frac{b(\lambda;x,t)}{a(\lambda;x,t)},
	\quad
	\hat{r}(\lambda;x,t)
	=
	\frac{\hat{b}(\lambda;x,t)}{\hat{a}(\lambda;x,t)},
	\qquad
\end{eqnarray}
so that on one side we have an incident and a reflected wave whereas on the other side we have a transmitted wave.

If we define a Wronskian-like function to be the determinant between different solutions, we can compute a set of relations
which allows one to determine the expansion coefficients
\begin{eqnarray}
	W[\vec \psi_{1}^{(I)}(\lambda;x,t),\vec \psi_{2}^{(II)}(\lambda;x,t)]
	=
	+a(\lambda;x,t),
	\qquad
	W[\vec \psi_{2}^{(I)}(\lambda;x,t),\vec \psi_{1}^{(II)}(\lambda;x,t)]
	=
	-\hat{a}(\lambda;x,t),
	\nonumber
	\\[8pt]
	\nonumber
	W[\vec \psi_{1}^{(I)}(\lambda;x,t),\vec \psi_{1}^{(II)}(\lambda;x,t)]
	=
	-b(\lambda;x,t),
	\qquad
	W[\vec \psi_{2}^{(I)}(\lambda;x,t),\vec \psi_{2}^{(II)}(\lambda;x,t)]
	=
	+\hat{b}(\lambda;x,t).
\end{eqnarray}
Since the object
\begin{eqnarray}
	W\left[
	\vec{\psi}_{1}(\lambda;x,t),
	\vec{\psi}_{2}(\lambda;x,t)
	\right] 
	=
	\psi_{11}(\lambda;x,t) \, \psi_{22}(\lambda;x,t)
	-
	\psi_{21}(\lambda;x,t) \, \psi_{12}(\lambda;x,t)
	\qquad\qquad
	\nonumber
\end{eqnarray}
satisfies
$
	\frac{d}{dx} W(\lambda;x,t)
	=
	\frac{d}{dx}  
	\big( \psi_{11} \, \psi_{22} - 
	\psi_{21} \, \psi_{12} \big)
	~=~
	0
$,
due to the equations of motion, one can see that these objects are x-independent and therefore we can write them as
\begin{eqnarray}
	a(\lambda;t)
	=
	+
	W[\vec \psi_{1}^{(I)}(\lambda;x,t),\vec \psi_{2}^{(II)}(\lambda;x,t)],
	\qquad
	\hat{a}(\lambda;t)
	=
	+
	W[\vec \psi_{1}^{(II)}(\lambda;x,t),\vec \psi_{2}^{(I)}(\lambda;x,t)],
	\qquad
	\\[8pt]
	b(\lambda;t)
	=
	+
	W[\vec \psi_{1}^{(II)}(\lambda;x,t),\vec \psi_{1}^{(I)}(\lambda;x,t)],
	\qquad
	\hat{b}(\lambda;t)
	=
	+
	W[\vec \psi_{2}^{(I)}(\lambda;x,t),\vec \psi_{2}^{(II)}(\lambda;x,t)].
	\qquad
\end{eqnarray}
Thus, in the case where the scattering is reflectionless, the wave passes through with matching boundary conditions
$\vec \psi_{1}^{(II)}(\lambda;x,t)=\vec \psi_{1}^{(I)}(\lambda;x,t)$ and $\vec \psi_{2}^{(II)}(\lambda;x,t)=\vec \psi_{2}^{(I)}(\lambda;x,t)$, so
\begin{eqnarray}
	a(\lambda;t)
	~=~
	1,
	\qquad
	\hat{a}(\lambda;t)
	~=~
	1,
	\qquad
	b(\lambda;t)
	~=~
	0,
	\qquad
	\hat{b}(\lambda;t)
	~=~
	0.
	\nonumber
\end{eqnarray}
In fact, when $b(\lambda;t)$ vanishes it means that the incoming wave is exactly the same as the outgoing wave. In other words, the incident and transmitted waves are the same, indicating that there is no reflection and this is the reflectionless scattering condition. 

Bound states, $\lambda_n$, are associated with the vanishing of the incoming wave so that only the transmitted and reflected waves are in place, so the discrete spectrum of the problem is given by the zeroes of the $a(\lambda_n,t)$, i.e. 
\begin{eqnarray}
	a(\lambda_n,t) &=& 0,
\end{eqnarray}
corresponding to simple poles of the reflection $r(\lambda;t)$ and transmission $\tau(\lambda;t)$ coefficients and thus,
\begin{eqnarray}
	\vec \psi_{1}^{(I)}(\lambda_n;x,t)
	&=&
	b(\lambda_n) \, \vec \psi_{2}^{(II)}(\lambda_n;x,t) 
	~=~
	c_n \, \vec \psi_{2}^{(II)}(\lambda_n;x,t).
\end{eqnarray}
In other words, the simple poles of the transmission and reflection coefficients, $\tau(\lambda;t)$ and $r(\lambda;t)$, respectively, correspond to bound states and can be found to be zeroes of the coefficient $a(\lambda;t)$.
If $a(\lambda;t)$ is an analytic function in a region of the complex plane, such as the upper plane, it can only have simple poles at
$\lambda_j = i \, \Lambda_j$, therefore at the upper-half plane.
Moreover, when $|\lambda| \to \infty$, the potential becomes negligible and the solution behaves as a plane wave, where indeed $a(\lambda;t) \to 1$ and $b(\lambda;t) \to 0$.
Overall, one may see that all the information about the scattering  is encoded in the reflection coefficient $r(\lambda;t) = \frac{b(\lambda;t)}{a(\lambda;t)}$.
The function $a(\lambda;t)$ is called the Jost function and constitutes the scattering data, together with the function $b(\lambda;t)$ and the numbers $c_n$ and $\lambda_n$.

\section{Time evolution of the spectral data in a quasi-integrable system}\label{secTimeEvol}\qquad

Thus, if one is concerned only with the in and out-states in a far past and a far future, respectively, there is the possibility of constructing a great deal of objects which are conserved in a broader sense. In order to do so, it is essential to investigate the behaviour of the spectral data, $a(\lambda;t)$ and $b(\lambda;t)$.
An important property of the spectral coefficients $a(\lambda)$ and $b(\lambda)$ is that the action-angle variables are constructed in terms of them.
This implies one can expand the action variable in powers of the spectral parameter $\lambda$, so that if it is constant, then one can generate infinitely many conserved charges. Relaxing this latter condition a bit, if the action variable changes slowly with time, keeping its final state equal to its initial state, then we have an infinite number of quantities whose values are asymptotically the same.

Considering the first Lax pair encountered, \eqref{standardLaxPairSGMt} and \eqref{standardLaxPairSGMx}, the  evolution of the scattering data can be done using the auxiliary linear equation 
$$\partial_{t} \vec{\psi}_j = A_{t} \vec{\psi}_j,$$ 
with solution of the form $\vec{\psi}_j = g \, \vec{\psi}_j^{(II)}$, as $x \to \infty$, so its time dependence is given by
\begin{eqnarray*}
	\begin{pmatrix}
		\partial_{t} g \, \psi_{1j}^{(II)}
		\\
		\partial_{t} g \, \psi_{2j}^{(II)} 
	\end{pmatrix}
	&=& 
	\begin{pmatrix}
		-i\Omega_{1}^{+} g \, \psi_{2j}^{(II)} 
		\\
		-i\Omega_{1}^{+} g \, \psi_{1j}^{(II)} 
	\end{pmatrix}
\end{eqnarray*}
with 
$
\Omega_1^\pm =
\frac{1}{4}  
\frac{\left(\lambda +\frac{1}{\lambda }\right)
	\mathcal{V}'(\phi_\pm)}{\sqrt{2} \sqrt{\mathcal{V}(\phi_\pm)}},
\Omega_2^\pm =
\frac{1}{4}\frac{\beta  \left(\lambda -\frac{1}{\lambda }\right)\sqrt{\mathcal{V}(\phi_\pm)}}{\sqrt{2}}
=0
$, 
from which one notes that
\begin{eqnarray*}
	\partial_{t} g
	&=&
	- i {\Omega_{1}^{+}} g.
\end{eqnarray*}
Regarding the behaviour as $x \to -\infty$, we perform the same analysis using (\ref{abexpansion1})
so that asymptotically the expressions
\begin{eqnarray*}
	\partial_{t}a(\lambda;x,t)
	&=&
	\pm i \, \big(  \Omega_{1}^{+} - \Omega_{1}^{-} \big) \, a(\lambda;x,t),
	\\[8pt]
	\partial_{t}b(\lambda;x,t)
	&=&
	\pm i \, \big(  \Omega_{1}^{+} + \Omega_{1}^{-} \big) \, b(\lambda;x,t),
\end{eqnarray*}
are valid.
As the vacua are degenerate, then $\mathcal{V}(\phi_+) = \mathcal{V}(\phi_-)$, so 
$\Omega_{1}^{+} = \Omega_{1}^{-} = \Omega_1$
and thus the equations above become
\begin{eqnarray}
	\partial_{t}a(\lambda;x,t)
	&=&
	0,
	\label{aconst1}
	\\[8pt]
	\partial_{t}b(\lambda;x,t)
	&=& \pm 2 i \,  \Omega_{1} \, b(\lambda;x,t).
\end{eqnarray}
We then see that the evolution of the scattering data is still simple in the case of this deformation. In particular, we deduce that $a(\lambda;t)=a(\lambda)$ is time-independent, thus being the generating function of the conservation laws. Moreover, by definition $a(\lambda_{n})=0$, therefore the zeroes, $\lambda_{n}$, are also time-independent.
Repeating the analysis for the evolution equation associated to the second gauge potentials, we obtain similar results.

\section{The study of specific quasi-integrable potentials}\label{secStudyPotential}\qquad

\subsection{A trigonometric deformation of the sine-Gordon model}\qquad

\begin{figure}[h!]
	\begin{center}
		\includegraphics[scale=0.7]{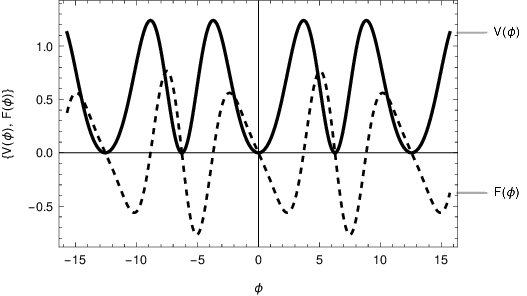}
		\caption{Potential and force fields, in continuous and dashed lines, respectively, associated to the model defined by (\ref{BZV}) with $\mu=1,\beta=1$ and $n=4$.}\label{BZVF}
	\end{center}
\end{figure}

The potential
\begin{eqnarray}
	\mathcal{V}(\phi) 
	&=& 
	\frac{32}{n^2} \frac{\mu^2}{\beta^2} 
	\tan^2{\left( \frac{\beta \phi} {4} \right)} 
	\left(1 - \sin^n{\left( \frac{\beta \phi}{4}\right)} \right)^2, \label{BZV}
\end{eqnarray}
defined in \cite{bazeia2008classical}, and already explored in \cite{ferreira2012concept}, is represented graphically in Figure \ref{BZVF}, alongside the associated force field,
which reduces to the sine-Gordon model for $n=2$ and vanishes for 
$
\phi_\pm
=
4N_{\pm}\frac{\pi}{\beta}
$, which implies that
$
	\mathcal{V}(\phi^\pm) = 0, 
	\mathcal{V}'(\phi^\pm) = 0
$ and
	$\frac{\mathcal{V}'(\phi_\pm)}{\sqrt{\mathcal{V}(\phi_\pm)}}= 2\sqrt{2} \frac{\mu}{n}
$.
From our previous discussion one can verify that indeed the scattering data has a simple time evolution
\begin{eqnarray}
	\partial_{t} a(\lambda;t) = 0, \quad \partial_{t} b(\lambda;t) = 2 i \Omega_1 b(\lambda;t),
\end{eqnarray}
with 
$
\Omega_1 = 
-\frac{\mu \cosh (\theta )}{n}
$ and $\lambda = e^\theta$.

\begin{figure}[h!]
	\begin{center}
		\includegraphics[scale=0.8]{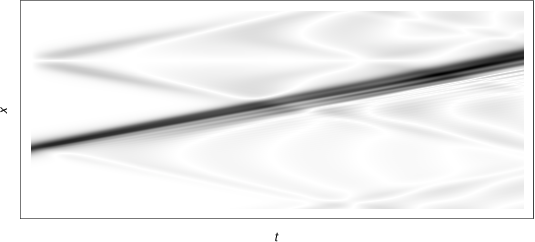}
		\quad
		\includegraphics[scale=0.8]{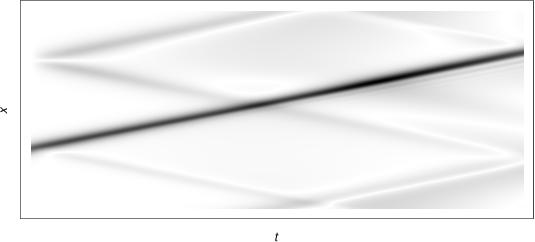}
		\caption{Time evolution, represented in a density plot form in terms of the $x$ and $t$ variables, for the field and momentum satisfying the dynamics given by (\ref{BZV}) with $\mu=1,\beta=1,n=4$ and initial condition given by two colliding solitonic waves.}\label{BZphi}
	\end{center}
\end{figure}

\begin{figure}[h!]
	\begin{center}
		\includegraphics[width=0.4\textwidth]{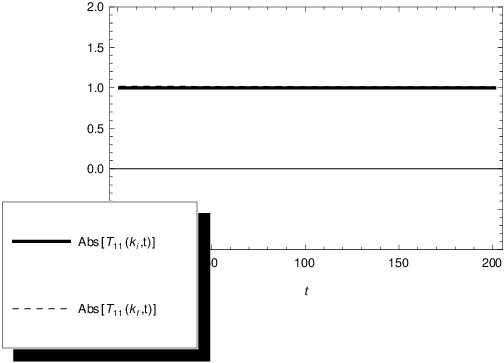}
		\qquad
		\includegraphics[width=0.4\textwidth]{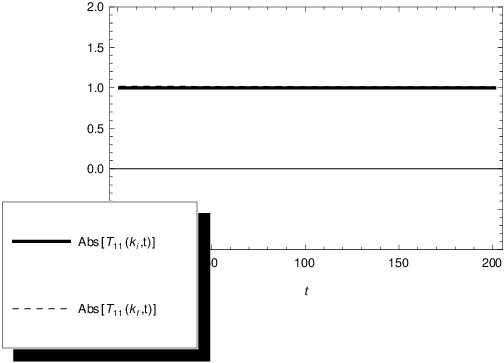}
		\caption{Time evolution of the absolute value $|a(\lambda,t)|$ and argument $Arg[a(\lambda,t)]$ associated to the scattering coefficient $a(\lambda,t)=T_{11}(\lambda,t)$ as a result of the dynamics in (\ref{BZV}) with $\mu=1,\beta=1,n=4$. The coinciding curves, represented by thick and dashed lines, represent different values of the spectral parameter.}\label{BZScatData}
	\end{center}
\end{figure}

In order to confirm the existence of a family of conserved quantities associated to the scattering data, or the entries of the transition matrix, $T(\lambda)$, we compute the numerical evolution for $n=4$ of a scattering between two solitonic solutions \cite{ferreira2012concept}. The time-dependence of the field and conjugate momentum are given by Figure \ref{BZphi}.
By computing the transition matrix one can plot the evolution of the element $a(\lambda,t)=T_{11}(\lambda,t)$ and the result is shown in Figure \ref{BZScatData}.
The scattering coefficient $a(\lambda,t)$ is seen to be real, with vanishing complex argument, and constant in time for the depicted interval, and two values of the spectral parameter, even during the interaction associated to the collision of the solitonic waves.
This is not a general property of this model but relies on the form of the initial condition and on space-time symmetry \cite{assis2016symmetry} of the scattering, in order to satisfy the necessary conditions for Equation (\ref{aconst1}) to be valid.
In principle, one could move on to the inverse scattering by using a Gel'fand-Levitan-Marchenko-like equation and try to construct the soliton solution. We leave this problem for a future investigation and, rather, focus on the conservation laws for other quasi-integrable models.
In what follows, we present a similar analysis for a variety of nonintegrable models.

\subsection{The double sine-Gordon model}\qquad

\begin{figure}[h!]
	\begin{center}
		\includegraphics[width=0.8\textwidth]{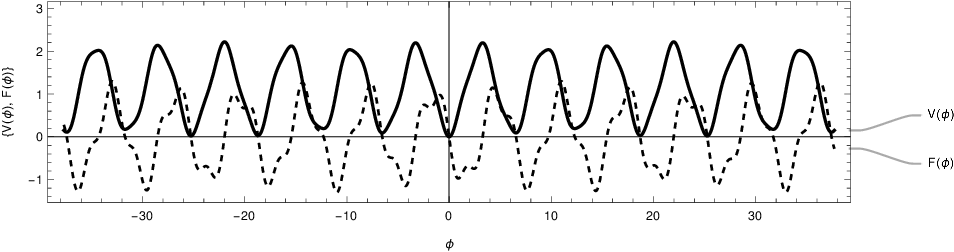}
		\caption{Potential and force fields, in continuous and dashed lines, respectively, associated to the model defined by (\ref{DSGV}) with $\mu=1, \nu=1,\beta=1,\gamma=e$.} \label{DSGVF}
	\end{center}
\end{figure}

\begin{figure}[h]
	\begin{center}
		\includegraphics[width=0.4\textwidth]{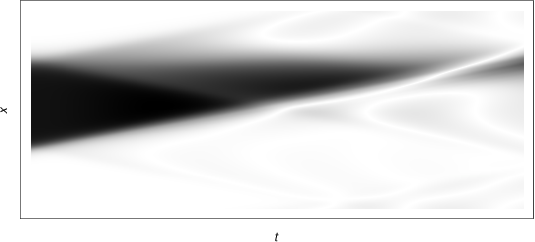}
		\quad
		\includegraphics[width=0.4\textwidth]{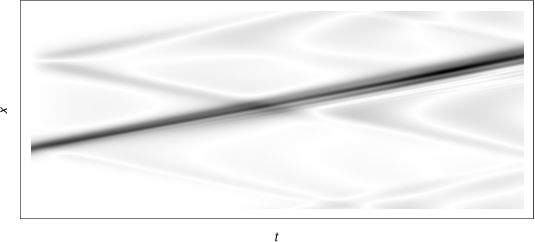}
		\caption{Time evolution, represented in a density plot form in $x$ and $t$ parameters, for the field and momentum satisfying the dynamics given by (\ref{DSGV}) with $\mu=1,\nu=1,\beta=1,\gamma=e$ and initial condition given by two colliding solitonic waves.}
		\label{DSGphi}
	\end{center}
\end{figure}

\begin{figure}[h]
	\begin{center}
		\includegraphics[width=0.4\textwidth]{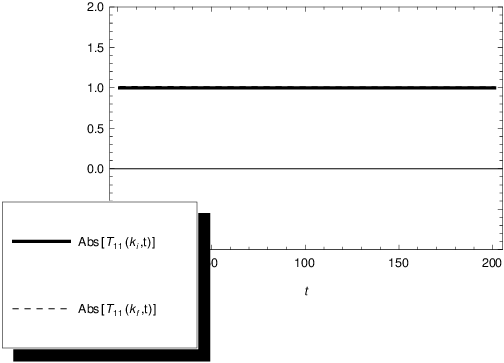}
		\qquad
		\includegraphics[width=0.4\textwidth]{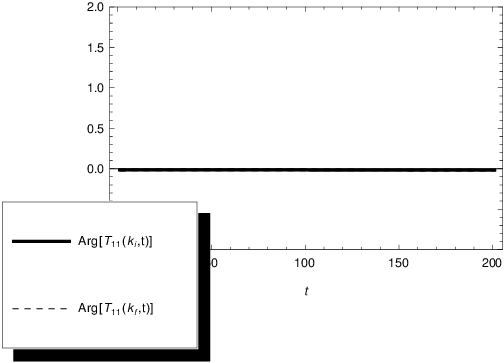}
		\caption{Time evolution of the absolute value $|a(\lambda,t)|$ and argument $Arg[a(\lambda,t)]$ associated to the scattering coefficient $a(\lambda,t)=T_{11}(\lambda,t)$ as a result of the dynamics in (\ref{DSGV}) with $\mu=1,\nu=1,\beta=1,\gamma=e$. The coinciding curves, represented by thick and dashed lines, represent different values of the spectral parameter.} \label{DSGScatData}
	\end{center}
\end{figure}

One familiar field theory that is closely related to the sine-Gordon model is the so called double sine-Gordon model, described by
\begin{eqnarray}
	\mathcal{V}(\phi)
	&=&
	\frac{\mu^2}{\beta^2} \left( 1 - \cos \beta\phi \right)
	+
	\frac{\nu^2}{\gamma^2} \left( 1 - \cos \gamma\phi \right)
	\label{DSGV}
\end{eqnarray}
and represented graphically in Figure \ref{DSGVF}.
Despite being closely related to the integrable sine-Gordon model, this other model is not integrable, that is, one cannot find a Lax pair representation for it.
In fact, the associated curvature does not vanish and the anomaly has the form
\begin{eqnarray}
	\frac{\Delta}{\frac{i \beta}{4}}
	=
	-
	\left( \Box \phi + \mathcal{V}'(\phi) \right) \, F_0
	+
	\nu^2 \left( 
	\frac{\beta^2}{\gamma^2} + \left(1-\frac{\beta^2}{\gamma^2}\right) \cos(\gamma\phi) 
	\right) 
	\big(
	\partial_t \phi -\partial_x \phi
	\big)\, F_1
	\qquad\qquad
\end{eqnarray}
Using as initial conditions two sine-Gordon solitons travelling in opposite directions represents an attempt to keep the curvature small and the following evolution, which can be seen in Figure \ref{DSGphi}. This
indicates again, but to a smaller degree than in the previous case, a symmetric behaviour with respect to the space and time variables.
As a result, the conservation property of the scattering data $a(\lambda,t)$ can be verified in Figure \ref{DSGScatData}.

\subsection{A deformed signum-Gordon model}\qquad

\begin{figure}[h!]
	\begin{center}
		\includegraphics[scale=0.6]{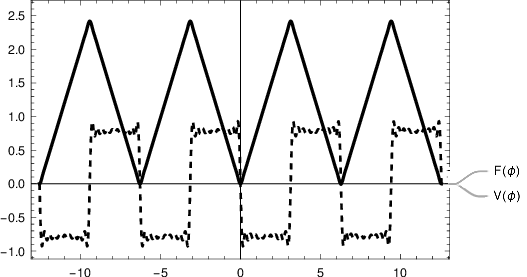}
		\caption{Potential and force fields, in continuous and dashed lines respectively, associated to the model defined by \eqref{sigGordo} with $\mu=1,\beta=1$.} \label{SiGV}
	\end{center}
\end{figure}

\begin{figure}[h]
	\begin{center}
		\includegraphics[width=0.37\textwidth]{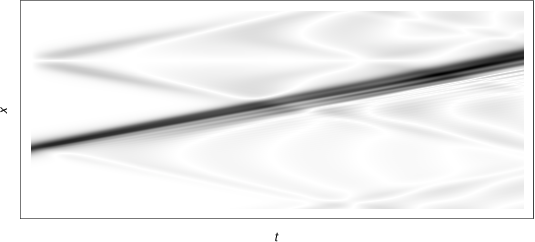}
		\quad
		\includegraphics[width=0.27\textwidth]{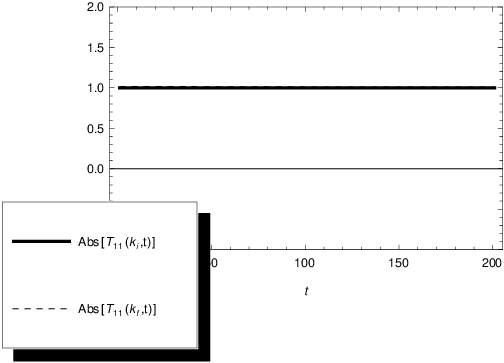}
		\quad
		\includegraphics[width=0.27\textwidth]{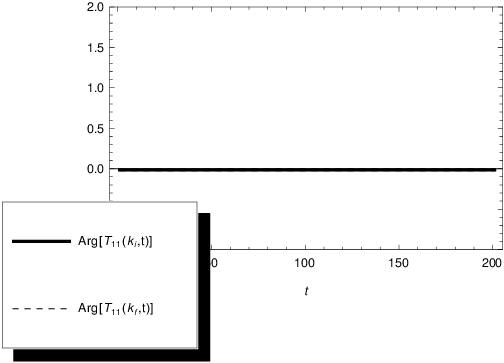}
		\caption{Time evolution associated to the dynamics of (\ref{SiGV}). On the left, a density plot of the field in space-time; in the middle, the absolute of the scattering data; on the right, the argument of the scattering data.}\label{SiGEv}
	\end{center}
\end{figure}

Another possibility, as a nonlinear Klein-Gordon equation, is the signum-Gordon model \cite{arodz2007signum,arodz2005field}. Here we implement a deformation of this model in the sense that we use a truncated Fourier expansion to generate the potential, 
\begin{eqnarray}\label{sigGordo}
	\mathcal{V}(\phi) 
	&=& 
	\sum_n
	\frac{1}{2n^2} \frac{\mu^2}{\beta^2} 
	\left(\cos(n\pi)-1\right)\left(\cos(n\beta\phi)-1\right),
\end{eqnarray}
represented in Figure \ref{SiGV}.
The time evolution and behaviour of the scattering data are presented in Figure \ref{SiGEv}
and represent no qualitative difference with respect to the previous examples\textcolor{red}{,} but reinforces the existence of conservation laws for a variety of quasi-integrable deformations.

\subsection{The $\phi^4$ theory}\qquad

\begin{figure}[h!]
	\begin{center}
		\includegraphics[scale=0.45]{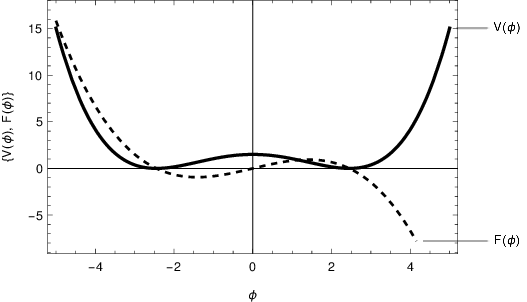}
		\quad
		\includegraphics[width=0.25\textwidth]{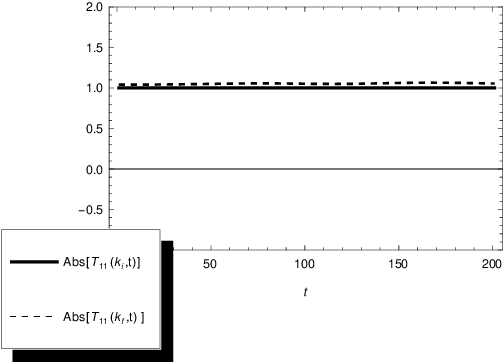}
		\quad
		\includegraphics[width=0.25\textwidth]{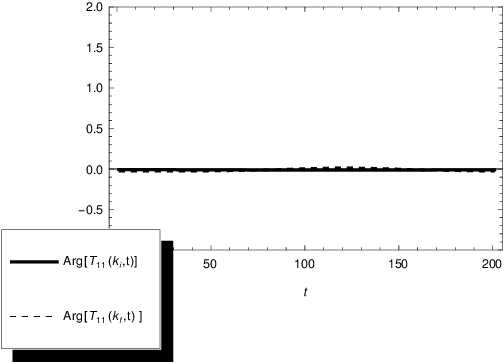}
		\caption{On the left, a representation of the potential and force profiles associated to (\ref{PhiV}); in the middle and on the right, time evolution of the quasi-integrable scattering coefficient.}\label{PhiEv}
	\end{center}
\end{figure}

Finally, we present a quasi-integrable formulation associated to the topological model known as the $\phi^4$ theory \cite{drell1976variational}, described by the potential
\begin{eqnarray}
	\mathcal{V}(\phi) 
	&=& 
	-\frac{\mu^2}{\beta^2} \left(
	\frac{\beta^2\phi^2}{2}-\frac{\beta^4\phi^4}{24}
	\right)\textcolor{red}{.}
	\label{PhiV}
\end{eqnarray}
As is well established, this system presents topological solitons and degenerate vacua, having two minima, located at $x= \pm x_0$, so that $\mathcal{V}(+x_0)=\mathcal{V}(-x_0)$ and $\mathcal{V}'(+x_0)=-\mathcal{V}'(-x_0)$.
Because of the topological properties, one does not usually attribute to quasi-integrability the interesting features of the model or assume there are asymptotically conserved charges.
Nonetheless, by carrying out our analyses, shown in Figure \ref{PhiEv}, one could conjecture that the existence of a solitonic solution associated to the $\phi^4$ model may be attributed to quasi-integrability.

\begin{figure}[h!]
	\begin{center}
		\includegraphics[scale=0.45]{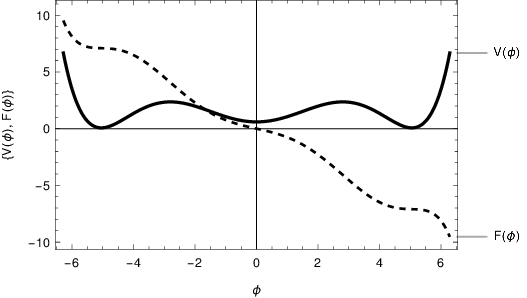}
		\quad
		\includegraphics[width=0.25\textwidth]{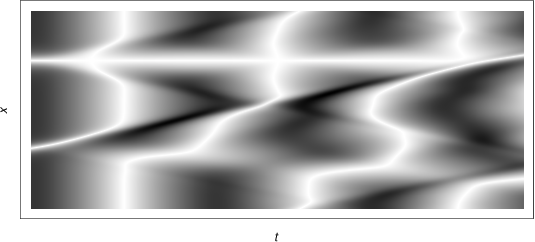}
		\quad
		\includegraphics[width=0.25\textwidth]{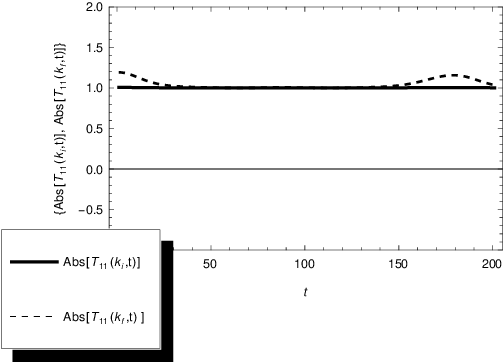}
		\caption{On the left, a representation of the potential and force profiles associated to (\ref{PPhiV}); in the middle, time evolution of the field; and on the right, time evolution of quasi-integrable scattering coefficient.}\label{PPhiEv}
	\end{center}
\end{figure}

Before finishing though, it is important not to leave the reader with the false impression that these conservation laws are abundant; instead, they are not so immediate to obtain. In this sense we show the results for a less fortunate example, of a $\phi^6$ model
\begin{eqnarray}
	\mathcal{V}(\phi) 
	&=& 
	+\frac{\mu^2}{\beta^2} \left(
	\frac{\beta^2\phi^2}{2}-\frac{\beta^4\phi^4}{24}+\frac{\beta^6\phi^6}{720}
	\right)
	\label{PPhiV}
\end{eqnarray}
presenting two degenerate vacua alongside a third vacuum which is very close to the other two, as can be seen in Figure \ref{PPhiEv}.
Here we note that the extra degeneracy induced by the vacuum at the origin affects the time evolution and, even for symmetric choices of initial conditions it is not easy to observe a conservation law associated to this model. In fact, as can be seen from Figure \ref{PPhiEv}, the scattering data is not time-independent.

\section{Concluding remarks}\label{secConclusion}\qquad

In this work we have presented a modification of the inverse scattering method for obtaining quasi-conservation laws associated to nonlinear Klein-Gordon equations.
For particular choices of initial conditions, symmetrical collisions of travelling solitonic waves, it was possible to represent the problems in terms of quasi-zero curvature conditions.
Then, the study of the scattering data for the associated linear problem showed that it was possible to have a time-independent family of conserved charges. 
Our numerical analysis made possible for one to verify that quasi-conservation laws can indeed be constructed for a broader class of problems, extending the reach of integrable tools to tackle nonintegrable deformation. The existence of time-independent scattering data is, however, far from general and we have presented an example where the method does not indicate quasi-integrability, which seems to lie somewhere in between usual integrability and topological soliton models.

\vspace{1cm}

\noindent {\bf Acknowledgements:}
P.E.G.A. is grateful for discussions with Luiz Agostinho Ferreira.
P.H.S.P. is supported by a CAPES/FAPEG grant, No. 202110267000487.

\bibliography{Bibliography}
\bibliographystyle{unsrt}

\end{document}